\documentclass[copyright,creativecommons]{eptcs}

\providecommand{\event}{PACO 2011}

\providecommand{\firstpage}{1}

\usepackage{amsmath,
	    amsfonts,
	    amstext,
	    amssymb,
	    mathrsfs}
\usepackage{latexsym}
\usepackage{proof}
\usepackage{breakurl}

\newcommand{\ultras}
	{\textsc{ULTraS}}

\newcommand{\mfun}[1]
	{\mathcal{M}_{#1}}

\def\ms#1{\null\ifmmode\mathord{\mathcode`-="702D\it #1\mathcode`\-="2200}%
	\else$\mathord{\mathcode`-="702D\it #1\mathcode`\-="2200}$\fi}

\newcommand{\cws}[2]
	{\\ \centerline{$#2$} \\[-#1pt]}

\newlength{\spacelen}

\newcommand{\calb}
        {\mathcal{B}}

\newcommand{\cald}
        {\mathcal{D}}

\newcommand{\cali}
        {\mathcal{I}}

\newcommand{\calo}
        {\mathcal{O}}

\newcommand{\cals}
        {\mathcal{S}}

\newcommand{\calu}
        {\mathcal{U}}

\newcommand{\bools}
	{\mathbb{B}}

\newcommand{\realns}
	{\mathbb{R}}

\newcommand{\procs}
	{\mathbb{P}}

\newcommand{\arrow}[2]
        {\, {\auxarrow\limits^{#1}}_{#2} \,}
\newcommand{\auxarrow}
{\mathop{\longrightarrow}}

\newcommand{\eqdef}
	{\buildrel \Delta \over =}

\newcommand{\sbis}[1]
	{\sim_{#1}}

\newcommand{\pco}[1]
	{\mathop{\Vert_{#1}}}

\newcommand{\fullbox}
	{{\mbox{}\nolinebreak\hfill{$\rule{2mm}{2mm}$}}}

\newtheorem{new_theorem}
	{Theorem}[section]

\newtheorem{new_definition}
	[new_theorem]{Definition}

\newtheorem{new_remark}
	[new_theorem]{Remark}

\newtheorem{new_example}
	[new_theorem]{Example}

\newtheorem{new_lemma}
	[new_theorem]{Lemma}

\newtheorem{new_proposition}
	[new_theorem]{Proposition}

\newtheorem{new_corollary}
	[new_theorem]{Corollary}

\newenvironment{definition}
	{\begin{new_definition}\rm}
	{\end{new_definition}}

\title{Uniform Labeled Transition Systems for Nondeterministic, Probabilistic, and Stochastic Process
Calculi}

\author{Marco Bernardo
\institute{Dipartimento di Scienze di Base e Fondamenti -- Universit\`a di Urbino -- Italy} 
\and Rocco De Nicola 
\institute{IMT -- Institute for Advanced Studies Lucca -- Italy \\
Dipartimento di Sistemi e Informatica -- Universit\`a di Firenze -- Italy}
\and Michele Loreti
\institute{Dipartimento di Sistemi e Informatica -- Universit\`a di Firenze -- Italy}}

\def\titlerunning{\ultras\ for Nondeterministic, Probabilistic, and Stochastic Process Calculi}
\def\authorrunning{M.\ Bernardo, R.\ De Nicola \& M.\ Loreti}

\begin{document}

\maketitle

\begin{abstract}

\noindent
Labeled transition systems are typically used to represent the behavior of nondeterministic processes, with
labeled transitions defining a one-step state-to-state reachability relation. This model has been recently
made more general by modifying the transition relation in such a way that it associates with any source
state and transition label a reachability distribution, i.e., a function mapping each possible target state
to a value of some domain that expresses the degree of one-step reachability of that target state. In this
extended abstract, we show how the resulting model, called \ultras\ from Uniform Labeled \textsc{Tra}nsition
System, can be naturally used to give semantics to a fully nondeterministic, a fully probabilistic, and a
fully stochastic variant of a CSP-like process language.

\end{abstract}


%
%
\section{Introduction}\label{sec:intro}
%
%

\newcommand{\amset}[1]{\mathscr{#1}}
\def\mtrans#1{\stackrel{#1}{\rightarrowtail}}

Process algebras are one of the most successful formalisms for modeling concurrent systems and proving their
properties such as correctness, liveness or safety. After their initial success in this respect, they have also been  
extended to deal with  properties related to performance and quality of service. 
Thus, process algebras have been enriched with quantitative notions of time and probabilities and 
integrated theories have been considered; for a comprehensive description of this approach, 
the reader is referred to \cite{ABC10}. 
Moreover, due to the growing interest in the analysis of shared-resource systems, stochastic variants 
of process algebras have also been proposed. 
The main aim being the integration of qualitative descriptions with those relative to performance 
in a single mathematical framework by building on th ecombination of labeled transition systems (LTS) 
and continuous-time Markov chains (CTMC). 

In~\cite{DLLM09a},  two of the authors of the present paper, together with D.~Latella and M.~Massink, proposed a variant of LTS,
namely {\em rate transition systems} (RTS), as a tool for providing semantics to some of the most
representative stochastic process languages. Within LTS, the transition relation describes the evolution of a system from one state to another as
determined by the execution of specific actions, thus it is a set of triples $(state,\ action,\ state)$. In
contrast, within RTS the transition relation $\mtrans{}$ associates with a given state $P$ and a given
transition label (action)~$a$ a function, say $\amset{P}$, mapping each term into a non-negative real
number. The transition $P\mtrans {a} \amset{P}$ has the following meaning: if $\amset{P}(Q) = v$ with $v
\not= 0$, then $Q$ is reachable from $P$ by executing $a$, the duration of such an execution being
exponentially distributed with rate $v$; if $\amset{P}(Q) = 0$, then $Q$ is not reachable from~$P$ via $a$.

RTSs have been used for providing a uniform semantic framework for modeling many of the different stochastic
process languages, facilitating reasoning about them, and throwing light on their similarities as well as on
their differences. In~\cite{DLLM09b}, we considered a limited, but representative, number of stochastic
process calculi and provided the RTS semantics for (fully) stochastic process languages both based
on the CSP-like, multipart interaction paradigm and on the CCS-like, two-ways interaction paradigm. Then,
in~\cite{DLLM11}, RTSs were extended by requiring that the domain of $\amset{P}$ be a generic semiring and
other variants of stochastic process algebras are studied, in particular it is shown that also languages, 
like IML~\cite{Her02}, that mix stochasticity and nondeterminism can be easily modeled.

In~\cite{BDL10}, we performed a further step in the direction of providing a uniform characterization of the
semantics of different process calculi and introduced a more general framework than RTS, which could be
instantiated to model not only stochastic process algebras but also classical process algebras, usually
modelled via LTS, and other quantitative variants of process algebras that would consider time,
probabilities, resources, etc.; we thus introduced \ultras{} (\emph{Uniform Labeled \textsc{Tra}nsition
Systems}). The transition relation of \ultras{} associates with a state and a given transition label a
function mapping each state into an element of a generic domain $D$. An \ultras{} transition $(s, a, \cald)$
is written $s \arrow{a}{} \cald$, with $\cald(s')$ being a $D$-value quantifying the degree of reachability
of $s'$ from $s$ via the execution of $a$ and $\cald(s') = \bot$ meaning that $s'$ is not reachable from $s$
via~$a$. By appropriately changing the domain $D$, different models of concurrent systems can be captured.
For example, if $D$ is the set $\bools$ consisting of the two Boolean values $true$ and $false$ we can
capture classical LTSs, while if $D$ is the set $\realns_{[0, 1]}$ we do capture probabilistic models, and
when $D$ is the set $\realns_{\ge 0}$ we do capture stochastically timed models.

Of course, modeling state transitions and their annotations is one of the key ingredients; however, one has
also to combine single transitions to obtain computations and find out ways for determining when two states
give rise to ``equivalent'' computation trees. To this aim, in~\cite{BDL10} we introduced the notions of
trace equivalence and bisimulation equivalence over \ultras{}. An important component of the equivalences
definition is a \emph{measure function} $\mfun{M}(s, \alpha, S')$ that computes the degree of multi-step
reachability of a set of target states $S'$ from a source state $s$ when performing computations labeled
with trace $\alpha$. For instance, to capture classical equivalences over nondeterministic systems, the
measure yields $\top$ if there exists a computation from $s$ to $S'$ labeled with $\alpha$ and $\bot$
otherwise. As another example, to capture probabilistic equivalences, the measure yields a value in
$\realns_{[0, 1]}$ that represents the probability of the set of computations labeled with $\alpha$ to reach
a state in $S'$ from $s$.

In this note, we put \ultras{} at work and use them to provide a uniform semantical description for a few
 (qualitative and quantitative) variants of a very simple process algebra. 
 For the sake of simplicity, we limit our attention to a purely
nondeterministic, a fully probabilistic, and a fully stochastic calculus, without allowing any interplay between
nondeterminism and quantitative aspects. In our view, the three (very compact) resulting sets of operational
rules give evidence of the expressive power of our approach and help in appreciating similarities and
differences among the three variants of the considered process algebra.

The rest of the paper is organized as follows. In Sect.~\ref{sec:ultras}, we recap the basic notions of
\ultras{} introduced in~\cite{BDL10} and define three different types of behavioral equivalences over them.
To the definition of trace and bisimulation equivalences already present in~\cite{BDL10}, we add the
definition of testing equivalence together with the set up of the necessary testing framework that we have introduced in~\cite{BDL11}.
In Sect.~\ref{sec:threecsp}, we show how \ultras{} can be used to provide the operational semantics of
classical CSP~\cite{BHR84} and of two of its probabilistic~\cite{Sei95,BBS95} and stochastic~\cite{Hil96}
variants. Finally, Sect.~\ref{sec:concl} reports on some future work.

%
%
\section{Uniform Labeled Transition Systems}\label{sec:ultras}
%
%

The behavior of sequential, concurrent, and distributed processes can be described by means of the so called
labeled transition system (LTS) model~\cite{Kel76}. It consists of a set of states, a set of transition
labels, and a transition relation. States correspond to the operational modes that processes can pass
through. Labels describe the activities that processes can perform internally or use to interact with the
environment. The transition relation defines process evolution as determined by the execution of specific
activities and is formalized as a \textit{state-to-state} reachability relation.

In this section, we recall from~\cite{BDL10} a generalization of the LTS model that aims at providing a
uniform framework that can be employed for defining and comparing the behavior of different types of
process. In the new model, named \ultras\ from Uniform Labeled \textsc{Tra}nsition System, the transition
relation associates with any source state and transition label a function mapping each possible target state
to an element of a domain~$D$. In other words, the state-to-state reachability relation typical of the LTS
model is replaced by a \textit{state-to-state-distribution} reachability relation. This is a consequence of
the fact that the concept of next state is generalized via a function that represents a one-step
reachability distribution, which expresses the degree of reachability from the current state of every
possible next state.

As shown in~\cite{BDL10}, by appropriately changing the domain $D$ we can capture different process models,
in particular quantitative ones like Markov chains~\cite{Ste94}. For example:

	\begin{itemize}

\item If $D$ is the support set $\bools = \{ \bot, \top \}$ of the Boolean algebra with the standard
conjunction ($\wedge$) and disjunction ($\vee$) operators, then we capture classical LTS models.

\item If $D = \realns_{[0, 1]}$, then we capture fully probabilistic models in the form of action-labeled
discrete-time Markov chains (ADTMC).

\item If $D = \realns_{\ge 0}$, then we capture fully stochastic models in the form of action-labeled
continuous-time Markov chains (ACTMC).

	\end{itemize}

%
\subsection{Definition of the Uniform Process Model}\label{sec:ultrasdef}
%

The definition of our uniform model is parameterized with respect to a complete partial order $(D,
\sqsubseteq)$ whose elements express the degree of \textit{one-step} reachability of a state. In the
following, we denote by~$\bot$ the $\sqsubseteq$-least element of $D$ and by $[S \rightarrow D]$ the set of
functions from a set $S$ to $D$, which is ranged over by~$\cald$.

	\begin{definition}

Let $(D, \sqsubseteq)$ be a complete partial order. A uniform labeled transition system on $(D,
\sqsubseteq)$, or $D$-\ultras\ for short, is a triple $\calu = (S, A, \! \arrow{}{} \!)$ where:

		\begin{itemize}

\item $S$ is an at most countable set of states.

\item $A$ is a countable set of transition-labeling actions.

\item $\! \arrow{}{} \! \subseteq S \times A \times [S \rightarrow D]$ is a transition relation.

		\end{itemize}

\noindent
We say that the $D$-\ultras{} $\calu$ is functional iff $\! \arrow{}{} \!$ is a function from $S \times A$
to $[S \rightarrow D]$.
\fullbox

	\end{definition}

Every transition $(s, a, \cald)$ is written $s \arrow{a}{} \cald$, with $\cald(s')$ being a $D$-value
quantifying the degree of reachability of $s'$ from $s$ via the execution of $a$ and $\cald(s') = \bot$
meaning that $s'$ is not reachable from $s$ via~$a$. When considering a functional \ultras, we will often
write $\cald_{s, a}(s')$ to denote the same $D$-value.

%
\subsection{Behavioral Equivalences for the \ultras\ Model}\label{sec:ultrasequiv}
%

LTS-based models come equipped with equivalences through which it is possible to compare processes on the
basis of their behavior and reduce the state space of a process before analyzing its properties. These
behavioral equivalences result in a linear-time/branching-time spectrum~\cite{Gla01,JS90,BKHW05,ABC10}
including several variants of three major approaches: bisimulation~\cite{HM85}, trace~\cite{BHR84}, and
testing~\cite{DH84}. We now recall how bisimulation, trace, and testing equivalences can be uniformly
defined over the \ultras\ model. 
Their definition
is parameterized
with respect to a measure function that expresses the degree of \textit{multi-step} reachability of a set of
states. Similar to the one-step reachability encoded within an \ultras, in which we consider individual
actions, multi-step reachability relies on sequences of actions commonly called traces, which are the
observable effects of the computations performed by an \ultras.

	\begin{definition}

Let $A$ be a countable set of transition-labeling actions. A trace $\alpha$ is an element of $A^{*}$, where
$\alpha = \varepsilon$ denotes the empty trace.
\fullbox

	\end{definition}

	\begin{definition}

Let $\calu = (S, A, \! \arrow{}{} \!)$ be a $D$-\ultras\ and $(M, \oplus, \otimes)$ be a lattice. An
$M$-measure function for $\calu$ is a function $\mfun{M} : S \times A^{*} \times 2^{S} \rightarrow M$.
\fullbox

	\end{definition}

Note that different measure functions can induce different variants of a behavioral equivalence on the same
$D$-\ultras\ depending on the support set and the operations of $(M, \oplus, \otimes)$. Although $D$ and~$M$
may be the same support set, this is not necessarily the case: while $D$-values are related to one-step
reachability, $M$-values -- especially those of the form $\mfun{M}(s, \alpha, S')$ -- are computed on the
basis of \linebreak $D$-values to quantify multi-step reachability.

\subsubsection{Trace Equivalence}

Trace equivalence is straightforward: two states are trace equivalent if every trace has the same measure
with respect to the entire set of states when starting from those two states.

	\begin{definition}

Let $\calu = (S, A, \arrow{}{} \!)$ be a $D$-\ultras\ and $\mfun{M}$ be an $M$-measure function for $\calu$.
\linebreak We say that $s_{1}, s_{2} \in S$ are $\mfun{M}$-trace equivalent, written $s_{1} \sbis{{\rm Tr},
\mfun{M}} s_{2}$, iff for all traces $\alpha \in A^{*}$:
\cws{10}{\mfun{M}(s_{1}, \alpha, S) \: = \: \mfun{M}(s_{2}, \alpha, S)}
\fullbox

	\end{definition}

\subsubsection{Bisimulation Equivalence}

While trace equivalence simply compares any two states without taking into account the states reached at the
end of the trace, bisimulation equivalence also poses constraints on the reached states.

	\begin{definition}

Let $\calu = (S, A, \! \arrow{}{} \!)$ be a $D$-\ultras\ and $\mfun{M}$ be an $M$-measure function for
$\calu$. An equivalence relation $\calb$ over $S$ is an $\mfun{M}$-bisimulation iff, whenever $(s_{1},
s_{2}) \in \calb$, then for all traces $\alpha \in A^{*}$ and equivalence classes $C \in S / \calb$:
\cws{0}{\mfun{M}(s_{1}, \alpha, C) \: = \: \mfun{M}(s_{2}, \alpha, C)}
We say that $s_{1}, s_{2} \in S$ are $\mfun{M}$-bisimilar, written $s_{1} \sbis{{\rm B}, \mfun{M}} s_{2}$,
iff there exists an $\mfun{M}$-bisimulation $\calb$ over~$S$ such that $(s_{1}, s_{2}) \in \calb$.
\fullbox

	\end{definition}

\subsubsection{Testing Equivalence}

The definition of testing equivalence requires the formalization of the notion of test and the consideration
of configurations rather than simple states. A test specifies which actions of a process are permitted at
each step and can be expressed as some suitable \ultras\ that includes a success state, which is used to determine which ones are the 
successful computations.

	\begin{definition}

Let $(D, \sqsubseteq)$ be a complete partial order. A $D$-observation system is a $D$-\ultras\ \linebreak
$\calo = (O, A, \! \arrow{}{} \!)$ where $O$ contains a distinguished success state denoted by $\omega$ such
that, whenever $\omega \arrow{a}{} \cald$, then $\cald(o) = \bot$ for all $o \in O$. We say that a
computation of $\calo$ is successful iff its length is finite and its last state is $\omega$.
\fullbox

	\end{definition}

A $D$-\ultras\ can be tested only through a $D$-observation system by running them in parallel and enforcing
synchronization on any action. The states of the resulting $D$-\ultras\ are called configurations and are
pairs each formed by a state of the $D$-\ultras\ under test and a state of the $D$-observation system. A
configuration can evolve to a new configuration only through the synchronization of two transitions --
leaving the two states constituting the configuration -- that are labeled with the same action and reach at
least one state, i.e., two identically labeled transitions whose target functions are not identically equal
to~$\bot$.

For each such pair of synchronizing transitions, the target function of the resulting transition is obtained
from the two original target functions by means of some $D$-valued function $\delta$, which computes the
degree of one-step reachability of every possible target configuration. Since $\bot$ represents
unreachability, the only constraint on $\delta$ is that it is $\bot$-preserving, i.e., that it yields $\bot$
iff at least one of its arguments is~$\bot$. As a consequence of this constraint, in the case of
nondeterministic processes $\delta$ boils down to logical conjunction, whereas several alternative options
are available in the case of probabilistic and stochastic processes.

	\begin{definition}

Let $\calu = (S, A, \! \arrow{}{\calu} \!)$ be a $D$-\ultras, $\calo = (O, A, \! \arrow{}{\calo} \!)$ be a
$D$-observation system, and $\delta$ be a $\bot$-preserving $D$-valued function. The interaction system of
$\calu$ and $\calo$ with respect to $\delta$ is the $D$-\ultras\ $\cali^{\delta}(\calu, \calo) = (S \times
O, A, \! \arrow{}{} \!)$ where:

		\begin{itemize}

\item Every element $(s, o) \in S \times O$ is called a configuration and is said to be successful iff $o =
\omega$. \linebreak We denote by $\cals^{\delta}(\calu, \calo)$ the set of successful configurations of
$\cali^{\delta}(\calu, \calo)$.

\item The transition relation $\! \arrow{}{} \! \subseteq (S \times O) \times A \times [(S \times O)
\rightarrow D]$ is such that $(s, o) \arrow{a}{} \cald$ iff \linebreak $s \arrow{a}{\calu} \cald_{1}$ and $o
\arrow{a}{\calo} \cald_{2}$ with $\cald(s', o')$ being obtained from $\cald_{1}(s')$ and $\cald_{2}(o')$ by
applying $\delta$. We say that a computation of $\cali^{\delta}(\calu, \calo)$ is successful iff its length
is finite and its last configuration is successful.
\fullbox

		\end{itemize}

	\end{definition}

	\begin{definition}

Let $\calu = (S, A, \! \arrow{}{\calu} \!)$ be a $D$-\ultras, $\mfun{M}$ be an $M$-measure function for
$\calu$, $\delta$ be a $\bot$-preserving $D$-valued function, and $\calo = (O, A, \! \arrow{}{\calo} \!)$ be
a $D$-observation system. The extension of $\mfun{M}$ to $\cali^{\delta}(\calu, \calo)$ is the function
$\mfun{M}^{\delta, \calo} : (S \times O) \times A^{*} \times 2^{S \times O} \rightarrow M$ whose definition
is obtained from that of $\mfun{M}$ by replacing states and transitions of $\calu$ with configurations and
transitions of $\cali^{\delta}(\calu, \calo)$.
\fullbox

	\end{definition}

	\begin{definition}

Let $\calu = (S, A, \! \arrow{}{\calu} \!)$ be a $D$-\ultras, $\mfun{M}$ be an $M$-measure function for
$\calu$, and $\delta$ be a $\bot$-preserving $D$-valued function. We say that $s_{1}, s_{2} \in S$ are
$\mfun{M}^{\delta}$-testing equivalent, written $s_{1} \sbis{{\rm T}, \mfun{M}^{\delta}}~s_{2}$, iff for all
$D$-observation systems $\calo = (O, A, \! \arrow{}{\calo} \!)$ with initial state $o \in O$ and for all
traces $\alpha \in A^{*}$:
\cws{10}{\mfun{M}^{\delta, \calo}((s_{1}, o), \alpha, \cals^{\delta}(\calu, \calo)) \: = \:
\mfun{M}^{\delta, \calo}((s_{2}, o), \alpha, \cals^{\delta}(\calu, \calo))}
\fullbox

	\end{definition}

%
%
\section{\ultras\ in Use: Three Experiments with CSP}\label{sec:threecsp}
%
%

\newcommand{\cspp}{\procs_{\rm CSP}}
\newcommand{\pcspp}{\procs_{\rm PCSP}}
\newcommand{\pepap}{\procs_{\rm PEPA}}
\newcommand{\rname}[1]{\mbox{\textsc{#1}}}
\newcommand{\total}{\oplus}

In this section, we show that the \ultras{} formalism can be used for providing operational models of
different kinds of process algebra. In particular, we will see how operational semantics of the language of
Communicating Sequential Processes (CSP)~\cite{BHR84} and two of its variants, which respectively extend the
calculus with probabilistic binary operators and exponentially timed actions, can be easily described within
the \ultras{} model by appropriately instantiating the domain $D$.

First, we introduce the syntax of the nondeterministic language and its operational semantics in terms of
\ultras. For the sake of simplicity, we only consider a kernel of CSP and omit some operators, like hiding
and renaming, because their treatment would add very little to the message we wish to convey. Then, we focus
on the probabilistic and stochastic variants of the kernel of CSP by exhibiting a suitable \ultras-based
operational semantics for each of them.

%
\subsection{$\bools$-\ultras\ Semantics for a Kernel of CSP}\label{sec:csp}
%

In CSP, systems are described as interactions of components that may engage in activities. Components
reflect the behavior of the important parts of a system, while activities capture the actions that the
components perform. The choice among the activities that are enabled in each system state is
nondeterministic.

Let $A$ be a countable set of activities. We denote by $\cspp$ the set of process terms defined according to
the following grammar:
\[\begin{array}{|c|}
\hline
\\[-0.3cm]
P \: ::= \: a . P \mid P + P \mid P \pco{L} P \mid B \\[0.1cm]
\hline
\end{array}\]
where $a \in A$, $L \subseteq A$, and $B$ is a behavioral constant defined by an appropriate equation of the
form $A \eqdef P$ for some process term $P$ in which constants occur only guarded in $P$, i.e., inside the
scope of an action prefix. Component $a . P$ models a process that performs activity $a$ and then behaves
like $P$. Component $P_{1} + P_{2}$ models a process that may behave either as~$P_{1}$ or as~$P_{2}$. The
operator $P_{1} \pco{L} P_{2}$ models instead the parallel execution of $P_{1}$ and $P_{2}$, which
synchronize (or cooperate) on every activity in $L$ and proceed independently on every activity not in~$L$.
The behavior of constant $B$ is the same as that of the process term $P$ on the right-hand side of its
defining equation.

The semantics for the considered kernel of CSP can be described in terms of the following functional
$\bools$-\ultras:
\cws{0}{(\cspp, A, \! \arrow{}{} \!)}
whose transition relation $\arrow{}{}$ is defined in Table~\ref{tab:cspopsem}. Given a transition $P
\arrow{a}{} \cald$, intuitively we have that $\cald(Q) = \top$ means that $Q$ is reachable from $P$ via an
$a$-transition, while $\cald(Q) = \bot$ means that it is not possible to reach $Q$ from $P$ by executing
$a$. 

\begin{table}[tbp]
\[
\begin{array}[center]{|c|}
\hline
\infer[\mbox{\textsc{Act}}]{
a . P \arrow{a}{} [P \mapsto \top]}{
}
\qquad
\infer[\rname{$\emptyset$-Act}]{
a . P \arrow{b}{} [ \, ]
}{
b \not= a
}
\qquad
\infer[\rname{Call}]{
B \arrow{a}{} \amset{D}
}{
B \eqdef P &
P \arrow{a}{} \amset{D}
}
\\[0.5cm]
\infer[\rname{Sum}]{
P_{1} + P_{2} \arrow{a}{} \amset{D}_1 \vee \amset{D}_2
}{
P_{1} \arrow{a}{} \amset{D}_1 &
P_{2} \arrow{a}{} \amset{D}_2
}
\\[0.5cm]
\infer[\rname{Coop}]{
P_{1} \pco{L} P_{2} \arrow{a}{} \amset{D}_1 \pco{L} \amset{D}_2
}{
P_{1} \arrow{a}{} \amset{D}_1 &
P_{2} \arrow{a}{} \amset{D}_2 &
a \in L
}
\\[0.5cm]
\infer[\rname{Int}]{
P_{1} \pco{L} P_{2} \arrow{a}{} (\amset{D}_1 \pco{L} P_{2}) \lor (P_{1} \pco{L} \amset{D}_2)
}{
P_{1} \arrow{a}{} \amset{D}_1 &
P_{2} \arrow{a}{} \amset{D}_2 &
a \notin L
}
\\[0.2cm]
\hline
\end{array}
\]
\caption{\ultras-based operational semantic rules for CSP}
\label{tab:cspopsem}
\end{table}

Rule \rname{Act} states that $a . P$ evolves via $a$ to $[P \mapsto \top]$, with the latter being the
function associating $\top$ with $P$ and $\bot$ with all the other process terms. On the contrary,
\rname{$\emptyset$-Act} establishes that no state is reachable from $a . P$ by performing any action $b
\not= a$. This is formalized by letting $a . P$ evolve via $b \not= a$ to $[ \, ]$, the function associating
$\bot$ with each process term. Rule \rname{Sum} describes nondeterministic choice: the states reachable from
$P_{1} + P_{2}$ via $a$ are all those that can be reached either by $P_{1}$ or by $P_{2}$. Indeed,
$\amset{D}_1 \vee \amset{D}_2$ denotes the function $\amset{D}$ such that $\amset{D}(Q) = \amset{D}_1(Q)
\vee \amset{D}_2(Q)$ for all process terms $Q$.

Rules \rname{Coop} and \rname{Int} govern parallel composition. Rule \rname{Coop} is used for computing the
next-state function when a synchronization between $P_{1}$ and $P_{2}$ occurs. Whenever $P_{1} \arrow{a}{}
\amset{D}_1$ and $P_{2} \arrow{a}{} \amset{D}_2$ with $a \in L$, then $P_{1} \pco{L} P_{2}$ evolves via $a$
to $\amset{D}_1 \pco{L} \amset{D}_2$, where $(\amset{D}_1 \pco{L} \amset{D}_2)(Q)$ is $\amset{D}_1(Q_{1})
\wedge \amset{D}_2(Q_{2})$ if $Q = Q_{1} \pco{L} Q_{2}$ and $\bot$ otherwise. Rule \rname{Int} deals with
$a \notin L$. In that case, if $P_{1} \arrow{a}{} \amset{D}_1$ and $P_{2} \arrow{a}{} \amset{D}_2$, then
$P_{1} \pco{L} P_{2}$ evolves via $a$ to $(\amset{D}_1 \pco{L} P_{2}) \lor (P_{1} \pco{L} \amset{D}_2)$,
where $\amset{D}_1 \pco{L} P_{2}$ (resp.\ $P_{1} \pco{L} \amset{D}_2$) denotes the function $\amset{D}$
such that $\amset{D}(Q)$ is $\amset{D}_1(P'_{1})$ (resp.\ $\amset{D}_2(P'_{2})$) if $Q = P'_{1} \pco{L}
P_{2}$ (resp. $Q = P_{1} \pco{L} P'_{2}$) and $\bot$ otherwise.

%
\subsection{$\realns_{[0, 1]}$-\ultras\ Semantics for PCSP}\label{sec:pcsp}
%

We now consider a probabilistic variant of CSP that we call PCSP. While in CSP the next action to execute is
selected nondeterministically, in PCSP it is selected according to some discrete probability distribution
that can be different from state to state. Taking inspiration from~\cite{Sei95,BBS95}, the probabilistic
calculus PCSP is obtained from CSP by decorating the alternative and parallel composition operators with a
probability value $p \in \realns_{[0, 1]}$.

We denote by $\pcspp$ the set of process terms defined according to the following grammar:
\[\begin{array}{|c|}
\hline
\\[-0.3cm]
P \: ::= \: a . P \mid P +_{p} P \mid P \pco{L}^{p} P \mid B \\[0.1cm]
\hline
\end{array}\]
Component $P_{1} +_{p} P_{2}$ models a process that, after performing an action, behaves as the continuation
of $P_{1}$ with probability $p$ or the continuation of $P_{2}$ with probability $1 - p$. Similarly, in
$P_{1} \pco{L}^{p} P_{2}$ the value $p$ is used to regulate the interleaving of $P_{1}$ and $P_{2}$.

The semantics for PCSP can be described in terms of the following functional $\realns_{[0, 1]}$-\ultras:
\cws{0}{(\pcspp, A, \! \arrow{}{} \!)}
whose transition relation $\arrow{}{}$ is defined in Table~\ref{tab:pcspopsem}. Given a transition $P
\arrow{a}{} \cald$, intuitively we have that $\cald(Q) > 0$ means that $Q$ is reachable from $P$ via an
$a$-transition with probability $\cald(Q)$, while $\cald(Q) = 0$ means that it is not possible to reach $Q$
from $P$ by executing $a$. Note that $\sum_{Q} \amset{D}(Q) \in \{ 0, 1 \}$.

\begin{table}[tbp]
\[
\begin{array}[center]{|c|}
\hline
\infer[\rname{Act}]{
a . P \arrow{a}{} [P \mapsto 1]}{
}
\qquad
\infer[\rname{$\emptyset$-Act}]{
a . P \arrow{b}{} [ \, ]
}{
b \not= a
}
\qquad
\infer[\rname{Call}]{
B \arrow{a}{} \amset{D}
}{
B \eqdef P &
P \arrow{a}{} \amset{D}
}
\\[0.5cm]
\infer[\rname{Sum}]{
P_{1} +_{p} P_{2} \arrow{a}{} \frac{p \cdot \oplus \amset{D}_1}{p \cdot \oplus \amset{D}_1 + (1 - p) \cdot
\oplus \amset{D}_2} \cdot \amset{D}_1 + \frac{(1 - p) \cdot \oplus \amset{D}_2}{p \cdot \oplus \amset{D}_1 +
(1 - p) \cdot \oplus \amset{D}_2} \cdot \amset{D}_2
}{
P_{1} \arrow{a}{} \amset{D}_1 &
P_{2} \arrow{a}{} \amset{D}_2
} 
\\[0.5cm]
\infer[\rname{Coop}]{
P_{1} \pco{L}^{p} P_{2} \arrow{a}{} \amset{D}_1 \pco{L} \amset{D}_2
}{
P_{1} \arrow{a}{} \amset{D}_1 &
P_{2} \arrow{a}{} \amset{D}_2 &
a \in L
}
\\[0.5cm]
\infer[\rname{Int}]{
P_{1} \pco{L}^{p} P_{2} \arrow{a}{} \frac{p \cdot \oplus \amset{D}_1}{p \cdot \oplus \amset{D}_1 + (1 - p)
\cdot \oplus{\amset{D}_2}} \cdot (\amset{D}_1 \pco{L} P_{2}) + \frac{(1 - p) \cdot \oplus \amset{D}_2}{p
\oplus \amset{D}_1 + (1 - p) \cdot \oplus \amset{D}_2} \cdot (P_{1} \pco{L} \amset{D}_2)
}{
P_{1} \arrow{a}{} \amset{D}_1 &
P_{2} \arrow{a}{} \amset{D}_2 &
a \not\in L
}
\\[0.2cm]
\hline
\end{array}
\]
\caption{\ultras-based operational semantic rules for PCSP}
\label{tab:pcspopsem}
\end{table}

The first three rules are identical to the first three rules of Table~\ref{tab:cspopsem}, with the
difference that $[P\mapsto 1]$ denotes the function associating $1$ with $P$ and $0$ with all the other
process terms, while $[ \, ]$ denotes the function associating $0$ with each process term. Rule \rname{Sum}
relies on the following notation:

	\begin{itemize}

\item $\amset{D}_1 + \amset{D}_2$ denotes the function $\amset{D}$ such that $\amset{D}(Q) = \amset{D}_1(Q)
+ \amset{D}_2(Q)$ for all process terms $Q$.

\item $\oplus \amset{D} = \sum_{Q} \amset{D}(Q)$.

\item $\frac{x}{y} \cdot \amset{D}$ denotes the function $\amset{D'}$ such that $\amset{D'}(Q) = \frac{x}{y}
\cdot \amset{D'}(Q)$ if $y \not= 0$ and $0$ otherwise.

	\end{itemize}

\noindent
This rule asserts that the states reachable from $P_{1} +_{p} P_{2}$ via $a$ are obtained by aggregating
according to~$p$ the probability distributions associated with $P_{1}$ and $P_{2}$ after $a$. When both
$P_{1}$ and $P_{2}$ can perform $a$, i.e., $P_{1} \arrow{a}{} \amset{D}_1$ and $P_{2} \arrow{a}{}
\amset{D}_2$ with $\amset{D}_1$ and $\amset{D}_2$ both different from $[ \, ]$, then $\oplus \amset{D}_{1} =
\oplus \amset{D}_{2} = 1$ and hence the aggregate probability distribution reduces to $p \cdot \amset{D}_1 +
(1 - p) \cdot \amset{D}_2$. In contrast, when $\amset{D}_1$ (resp.\ $\amset{D}_2$) is equal to $[ \, ]$,
then $\oplus \amset{D}_{1} = 0$ (resp.\ $\oplus \amset{D}_{2} = 0$) and hence the aggregate probability
distribution reduces to $\amset{D}_2$ (resp.\ $\amset{D}_1$).

Rules \rname{Coop} and \rname{Int} govern parallel composition. They are similar to the two corresponding
rules of Table~\ref{tab:cspopsem}, with the differences that (i) in the synchronization case $(\amset{D}_1
\pco{L} \amset{D}_2)(Q)$ is $\amset{D}_1(Q_{1}) \cdot \amset{D}_2(Q_{2})$ if $Q = Q_{1} \pco{L}^{p} Q_{2}$
and $0$ otherwise, while (ii) in the interleaving case a $\rname{Sum}$-like aggregation based on $p$ of the
probability distributions associated with $P_{1}$ and $P_{2}$ after $a$ comes into play.

%
\subsection{$\realns_{\ge 0}$-\ultras\ Semantics for PEPA}\label{sec:pepa}
%

Building on~\cite{DLLM09a,DLLM09b}, we finally consider a stochastically timed variant of CSP called
Performance Evaluation Process Algebra (PEPA)~\cite{Hil96}. In this calculus, every action is equipped with
a rate $\lambda \in \realns_{> 0}$ that uniquely characterizes the exponentially distributed random variable
quantifying the duration of the action itself (the expected duration is $1 / \lambda$). The choice among the
actions that are enabled in each state is governed by the race policy: the action to execute is the one that
samples the least duration. Therefore, (i) the sojourn time in each state is exponentially distributed with
rate given by the sum of the rates of the transitions departing from that state, (ii) the execution
probability of each transition is proportional to its rate, and (iii) the alternative and parallel
composition operators are implicitly probabilistic.

We denote by $\pepap$ the set of process terms defined according to the following grammar:
\[\begin{array}{|c|}
\hline
\\[-0.3cm]
P \: ::= \: (a, \lambda) . P \mid P + P \mid P \pco{L} P \mid B \\[0.1cm]
\hline
\end{array}\]
Component $(a, \lambda) . P$ models a process that can perform action $a$ at rate $\lambda$ and then behaves
like $P$.

\begin{table}[tbp]
\[
\begin{array}[center]{|c|}
\hline
\infer[\rname{Act}]{
(a, \lambda) . P \arrow{a}{} [P \mapsto \lambda]}{} \qquad
\infer[\rname{$\emptyset$-Act}]{
(a, \lambda). P \arrow{b}{} [ \, ]}{a \not= b} \qquad
\infer[\rname{Call}]{
B \arrow{a}{} \amset{D}
}{
B \eqdef P &
P \arrow{a}{} \amset{D}
}
\\[0.5cm]
\infer[\rname{Sum}]{
P_{1} + P_{2} \arrow{a}{} \amset{D}_1 + \amset{D}_2
}{
P_{1} \arrow{a}{} \amset{D}_1 &
P_{2} \arrow{a}{} \amset{D}_2
}
\\[0.5cm]
\infer[\rname{Coop}]{
P_{1} \pco{L} P_{2} \arrow{a}{} \frac{\min \{ \total \amset{D}_1, \total\amset{D}_2 \}}{\total \amset{D}_1
\cdot \total \amset{D}_2} \cdot (\amset{D}_1 \pco{L} \amset{D}_2)
}{
P_{1} \arrow{a}{} \amset{D}_1 &
P_{2} \arrow{a}{} \amset{D}_2 &
a \in L
}
\\[0.5cm]
\infer[\rname{Int}]{
P_{1} \pco{L} P_{2} \arrow{a}{} (\amset{D}_1 \pco{L} P_{2}) + (P_{1} \pco{L} \amset{D}_2)
}{
P_{1} \arrow{a}{} \amset{D}_1 &
P_{2} \arrow{a}{} \amset{D}_2 &
a \notin L
}
\\[0.2cm]
\hline
\end{array}
\]
\caption{\ultras-based operational semantic rules for PEPA}
\label{tab:pepaopsem}
\end{table}

The semantics for PEPA can be described in terms of the following functional $\realns_{\ge 0}$-\ultras:
\cws{0}{(\pepap, A, \! \arrow{}{} \!)}
whose transition relation $\arrow{}{}$ is defined in Table~\ref{tab:pepaopsem}. Given a transition $P
\arrow{a}{} \cald$, intuitively we have that $\cald(Q) > 0$ means that $Q$ is reachable from $P$ via an
$a$-transition at rate $\cald(Q)$, while $\cald(Q) = 0$ means that it is not possible to reach $Q$ from $P$
by executing $a$.

The rules of Table~\ref{tab:pepaopsem} are similar to those of Table~\ref{tab:pcspopsem}, with the
differences that (i) $[P \mapsto \lambda]$ denotes the function associating $\lambda$ with $P$ and $0$ with
all the other process terms, (ii) no normalization is needed in rules \rname{Sum} and \rname{Int} because
transition rates simply sum up due to the race policy, and (iii) the multiplicative factor in rule
\rname{Coop} is specific to the PEPA cooperation discipline based on the slowest component.

%
%
\section{Conclusions and Future Work}\label{sec:concl}
%
%

After recalling the \ultras\ model from~\cite{BDL10,BDL11}, in this paper we have extended the scope of the
work done in~\cite{DLLM09a,DLLM09b,DLLM11} by exhibiting the \ultras-based operational semantic rules for
CSP and two of its probabilistic and stochastically timed variants. These three experiments seem to indicate
that the \ultras\ model naturally lends itself to be used as a compact and uniform semantic framework for
different classes of process calculi.

With respect to future work, we plan to continue our experiments by using the \ultras\ model for describing
the operational semantics of other process description languages of nondeterministic, probabilistic, or
stochastic nature, as well as process calculi combining nondeterminism and probability or stochasticity.
This should help to assess the relative expressiveness of their operators and establish general properties
for the various languages. Moreover, the uniform characterization of the equivalences might help in evaluating 
and discerning among the many relations proposed in the literature. It would be, indeed, interesting to 
determine which of the existing relations can be obtained as instances of the general framework.

This study may also lead to the definition of a uniform process calculus with an \ultras-based operational
semantics and the development of uniform axiomatizations of bisimulation, trace, and testing equivalences.
From this calculus, it should be possible to retrieve the originally proposed calculi by varying the target
domain and the behavioral operators. We shall also consider  further options related to quantitative
aspects like including quantities within actions (\emph{integrated quantity approach}) or attaching them to traditional
operators or providing specific operators for them (\emph{orthogonal quantity approach}).

Finally, it would be interesting to see whether is is possible to build generic tools for supporting verifications that are based on 
the uniform model and have only to be instantiated to deal with the specific calculi.

\bigskip
\noindent
\textbf{Acknowledgment}: This work has been partially supported by the EU project ASCENS 257414.

\bibliographystyle{eptcs}
\bibliography{paco2011}

\begin{thebibliography}{10}
\providecommand{\bibitemdeclare}[2]{}
\providecommand{\urlprefix}{Available at }
\providecommand{\url}[1]{\texttt{#1}}
\providecommand{\href}[2]{\texttt{#2}}
\providecommand{\urlalt}[2]{\href{#1}{#2}}
\providecommand{\doi}[1]{doi:\urlalt{http://dx.doi.org/#1}{#1}}
\providecommand{\bibinfo}[2]{#2}

\bibitemdeclare{book}{ABC10}
\bibitem{ABC10}
\bibinfo{author}{A.~Aldini}, \bibinfo{author}{M.~Bernardo} \&
  \bibinfo{author}{F.~Corradini} (\bibinfo{year}{2010}):
  \emph{\bibinfo{title}{A Process Algebraic Approach to Software Architecture
  Design}}.
\newblock \bibinfo{publisher}{Springer}, \doi{10.1007/978-1-84800-223-4}.

\bibitemdeclare{article}{BBS95}
\bibitem{BBS95}
\bibinfo{author}{J.C.M. Baeten}, \bibinfo{author}{J.A. Bergstra} \&
  \bibinfo{author}{S.A. Smolka} (\bibinfo{year}{1995}):
  \emph{\bibinfo{title}{Axiomatizing Probabilistic Processes: {ACP} with
  Generative Probabilities}}.
\newblock {\sl \bibinfo{journal}{Information and Computation}}
  \bibinfo{volume}{121}, pp. \bibinfo{pages}{234--255},
  \doi{10.1006/inco.1995.1135}.

\bibitemdeclare{article}{BKHW05}
\bibitem{BKHW05}
\bibinfo{author}{C.~Baier}, \bibinfo{author}{J.-P. Katoen},
  \bibinfo{author}{H.~Hermanns} \& \bibinfo{author}{V.~Wolf}
  (\bibinfo{year}{2005}): \emph{\bibinfo{title}{Comparative Branching-Time
  Semantics for {M}arkov Chains}}.
\newblock {\sl \bibinfo{journal}{Information and Computation}}
  \bibinfo{volume}{200}, pp. \bibinfo{pages}{149--214},
  \doi{10.1016/j.ic.2005.03.001}.

\bibitemdeclare{inproceedings}{BDL10}
\bibitem{BDL10}
\bibinfo{author}{M.~Bernardo}, \bibinfo{author}{R.~{De Nicola}} \&
  \bibinfo{author}{M.~Loreti} (\bibinfo{year}{2010}):
  \emph{\bibinfo{title}{Uniform Labeled Transition Systems for
  Nondeterministic, Probabilistic, and Stochastic Processes}}.
\newblock In: {\sl \bibinfo{booktitle}{Proc.\ of the 5th Int.\ Symp.\ on
  Trustworthy Global Computing (TGC~2010)}}, {\sl \bibinfo{series}{LNCS}}
  \bibinfo{volume}{6084}, \bibinfo{publisher}{Springer}, pp.
  \bibinfo{pages}{35--56}, \doi{10.1007/978-3-642-15640-3}.

\bibitemdeclare{unpublished}{BDL11}
\bibitem{BDL11}
\bibinfo{author}{M.~Bernardo}, \bibinfo{author}{R.~{De Nicola}} \&
  \bibinfo{author}{M.~Loreti} (\bibinfo{year}{2011}): \emph{\bibinfo{title}{A
  Uniform Framework for Process Models and Behavioral Equivalences of
  Nondeterministic, Probabilistic, Stochastic, or Mixed Nature}}.
\newblock \bibinfo{note}{Submitted for journal publication}.

\bibitemdeclare{article}{BHR84}
\bibitem{BHR84}
\bibinfo{author}{S.D. Brookes}, \bibinfo{author}{C.A.R. Hoare} \&
  \bibinfo{author}{A.W. Roscoe} (\bibinfo{year}{1984}): \emph{\bibinfo{title}{A
  Theory of Communicating Sequential Processes}}.
\newblock {\sl \bibinfo{journal}{Journal of the ACM}} \bibinfo{volume}{31}, pp.
  \bibinfo{pages}{560--599}, \doi{10.1145/828.833}.

\bibitemdeclare{article}{DH84}
\bibitem{DH84}
\bibinfo{author}{R.~{De Nicola}} \& \bibinfo{author}{M.~Hennessy}
  (\bibinfo{year}{1984}): \emph{\bibinfo{title}{Testing Equivalences for
  Processes}}.
\newblock {\sl \bibinfo{journal}{Theoretical Computer Science}}
  \bibinfo{volume}{34}, pp. \bibinfo{pages}{83--133},
  \doi{10.1016/0304-3975(84)90113-0}.

\bibitemdeclare{inproceedings}{DLLM09b}
\bibitem{DLLM09b}
\bibinfo{author}{R.~{De Nicola}}, \bibinfo{author}{D.~Latella},
  \bibinfo{author}{M.~Loreti} \& \bibinfo{author}{M.~Massink}
  (\bibinfo{year}{2009}): \emph{\bibinfo{title}{On a Uniform Framework for the
  Definition of Stochastic Process Languages}}.
\newblock In: {\sl \bibinfo{booktitle}{Proc.\ of the 14th Int.\ Workshop on
  Formal Methods for Industrial Critical Systems (FMICS~2009)}}, {\sl
  \bibinfo{series}{LNCS}} \bibinfo{volume}{5825},
  \bibinfo{publisher}{Springer}, pp. \bibinfo{pages}{9--25},
  \doi{10.1007/978-3-642-04570-7\_2}.

\bibitemdeclare{inproceedings}{DLLM09a}
\bibitem{DLLM09a}
\bibinfo{author}{R.~{De Nicola}}, \bibinfo{author}{D.~Latella},
  \bibinfo{author}{M.~Loreti} \& \bibinfo{author}{M.~Massink}
  (\bibinfo{year}{2009}): \emph{\bibinfo{title}{Rate-Based Transition Systems
  for Stochastic Process Calculi}}.
\newblock In: {\sl \bibinfo{booktitle}{Proc.\ of the 36th Int.\ Coll.\ on
  Automata, Languages and Programming (ICALP~2009)}}, {\sl
  \bibinfo{series}{LNCS}} \bibinfo{volume}{5556},
  \bibinfo{publisher}{Springer}, pp. \bibinfo{pages}{435--446},
  \doi{10.1007/978-3-642-02930-1\_36}.

\bibitemdeclare{techreport}{DLLM11}
\bibitem{DLLM11}
\bibinfo{author}{R.~{De Nicola}}, \bibinfo{author}{D.~Latella},
  \bibinfo{author}{M.~Loreti} \& \bibinfo{author}{M.~Massink}
  (\bibinfo{year}{2011}): \emph{\bibinfo{title}{State to Function Labelled
  Transition Systems: A Uniform Framework for Defining Stochastic Process
  Calculi}}.
\newblock \bibinfo{type}{Technical Report}, \bibinfo{institution}{CNR-ISTI}.
\newblock
  \urlprefix\url{http://puma.isti.cnr.it/download.php?DocFile=2011-TR-012_0.pdf&idcode=2011-TR-012&authority=cnr.isti&collection=cnr.isti}.

\bibitemdeclare{inproceedings}{Gla01}
\bibitem{Gla01}
\bibinfo{author}{R.J. van Glabbeek} (\bibinfo{year}{2001}):
  \emph{\bibinfo{title}{The Linear Time -- Branching Time Spectrum~{I}}}.
\newblock In: {\sl \bibinfo{booktitle}{Handbook of Process Algebra}},
  \bibinfo{publisher}{Elsevier}, pp. \bibinfo{pages}{3--99},
  \doi{10.1007/BFb0039066}.

\bibitemdeclare{article}{HM85}
\bibitem{HM85}
\bibinfo{author}{M.~Hennessy} \& \bibinfo{author}{R.~Milner}
  (\bibinfo{year}{1985}): \emph{\bibinfo{title}{Algebraic Laws for
  Nondeterminism and Concurrency}}.
\newblock {\sl \bibinfo{journal}{Journal of the ACM}} \bibinfo{volume}{32}, pp.
  \bibinfo{pages}{137--162}, \doi{10.1145/2455.2460}.

\bibitemdeclare{book}{Her02}
\bibitem{Her02}
\bibinfo{author}{H.~Hermanns} (\bibinfo{year}{2002}):
  \emph{\bibinfo{title}{Interactive {M}arkov Chains}}.
\newblock \bibinfo{publisher}{Springer}, \doi{10.1007/3-540-45804-2}.
\newblock \bibinfo{note}{Volume 2428 of LNCS}.

\bibitemdeclare{book}{Hil96}
\bibitem{Hil96}
\bibinfo{author}{J.~Hillston} (\bibinfo{year}{1996}): \emph{\bibinfo{title}{A
  Compositional Approach to Performance Modelling}}.
\newblock \bibinfo{publisher}{Cambridge University Press},
  \doi{10.1017/CBO9780511569951}.

\bibitemdeclare{inproceedings}{JS90}
\bibitem{JS90}
\bibinfo{author}{C.-C. Jou} \& \bibinfo{author}{S.A. Smolka}
  (\bibinfo{year}{1990}): \emph{\bibinfo{title}{Equivalences, Congruences, and
  Complete Axiomatizations for Probabilistic Processes}}.
\newblock In: {\sl \bibinfo{booktitle}{Proc.\ of the 1st Int.\ Conf.\ on
  Concurrency Theory (CONCUR~1990)}}, {\sl \bibinfo{series}{LNCS}}
  \bibinfo{volume}{458}, \bibinfo{publisher}{Springer}, pp.
  \bibinfo{pages}{367--383}, \doi{10.1007/BFb0039071}.

\bibitemdeclare{article}{Kel76}
\bibitem{Kel76}
\bibinfo{author}{R.M. Keller} (\bibinfo{year}{1976}):
  \emph{\bibinfo{title}{Formal Verification of Parallel Programs}}.
\newblock {\sl \bibinfo{journal}{Communications of the ACM}}
  \bibinfo{volume}{19}, pp. \bibinfo{pages}{371--384},
  \doi{10.1145/360248.360251}.

\bibitemdeclare{article}{Sei95}
\bibitem{Sei95}
\bibinfo{author}{K.~Seidel} (\bibinfo{year}{1995}):
  \emph{\bibinfo{title}{Probabilistic Communicating Processes}}.
\newblock {\sl \bibinfo{journal}{Theoretical Computer Science}}
  \bibinfo{volume}{152}, pp. \bibinfo{pages}{219--249},
  \doi{10.1016/0304-3975(94)00286-0}.

\bibitemdeclare{book}{Ste94}
\bibitem{Ste94}
\bibinfo{author}{W.J. Stewart} (\bibinfo{year}{1994}):
  \emph{\bibinfo{title}{Introduction to the Numerical Solution of Markov
  Chains}}.
\newblock \bibinfo{publisher}{Princeton University Press}.
\newblock \urlprefix\url{http://press.princeton.edu/titles/5640.html}.

\end{thebibliography}

\end{document}